# Design of Highly Efficient Hybrid Si-Au Taper for Dielectric Strip Waveguide to Plasmonic Slot Waveguide Mode Converter

Chin-Ta Chen, Xiaochuan Xu, Amir Hosseini, Zeyu Pan, Harish Subbaraman, Xingyu Zhang, and Ray T. Chen

*Abstract*—In this paper, we design a dielectric-to-plasmonic slot waveguide mode converter based on the hybrid silicon-gold taper. The effects of mode matching, the effective index matching, and the metallic absorption loss on the conversion efficiency are studied. Consequently, a metallic taper-funnel coupler with an overall length of 1.7 $\mu$m is designed to achieve a very high conversion efficiency of 93.3% at 1550 nm. The configuration limitations for not allowing this mode converter to achieve a 100% conversion efficiency are also investigated. Such a high-efficiency converter can provide practical routes to realize ultracompact integrated circuits.

## I. Introduction

PLASMONIC devices enable linear and non-linear processing of light beyond the diffraction limit [1], and thus have drawn considerable attention in recent years. Several plasmonic waveguide structures such as metallic nanoparticle arrays and metallic nanowires have been proposed, but most of them only support highly confined modes near the surface plasmon frequency [2]. A metal-insulator-metal (MIM) waveguide structure supports nanometer-size modes over a wide wavelength range and therefore, is more intriguing for applications including MIM optical waveguides [2], [7]–[15], light sources [3], detectors [4], and modulators [5], [6], etc. However, MIM waveguides suffer from a high propagation loss ($>$1 dB/$\mu$m) [9], [12], [13], which limits the length of these waveguides to be less than 100 $\mu$m. Since low loss dielectric waveguides are capable of delivering photons over tens of centimeters long distance, it is a natural choice to use low loss dielectric waveguides to deliver photons into and out from MIM based plasmonic devices [7]. To implement such a system, high efficient dielectric-to-plasmonic slot waveguide couplers are of significant importance.

To realize low loss coupling, various coupler structures have been investigated, including a multisection taper with a theoretical efficiency of 93% according to two-dimensional (2-D) simulations (300-nm-wide silicon waveguide to 50-nm-wide MIM waveguide) [2], an air-gap coupler with 88% efficiency (300-nm-wide silicon waveguide to 40-nm-wide MIM waveguide) [11], a silicon slot waveguide with a theoretical efficiency of about 70% (silicon slot waveguide with a slot width of 120 nm to quasi-plasmonic slot waveguide) [12], a silicon-gold plasmonic coupler with a theoretical efficiency of 88% (450-nm-wide silicon waveguide to 200-nm-wide MIM waveguide) [13], a taper-funnel coupler with a theoretical efficiency of 33% (300-nm-wide silicon waveguide to 200-nm-wide MIM waveguide) [14], etc. Although these tapers show great promise, none of these could achieve zero loss conversion theoretically. An investigation on the root cause of the conversion loss would give some clues on how to increase the coupling efficiency even further, which is the focus of this paper.

The discussion is organized as follows. First, the proposed mode converter geometry and simulation configurations are described. The evolution of plasmonic slot waveguide mode within the hybrid silicon-gold taper is investigated in order to understand the root cause of conversion loss. Then, we study the hybrid silicon-gold taper from three aspects, specifically, mode matching, effective index matching, and metallic absorption loss. The configuration limitations for the hybrid silicon-gold taper that would possibly restrict the mode converter to achieve a near 100% conversion efficiency are also discussed. According to the above-mentioned discussions, a mode converter design providing conversion efficiency up to 93.3% at 1550 nm is presented.

## II. Design and Configuration of Mode Converter

The hybrid silicon-gold taper is shown in Fig. 1(a). Fig. 1(b)–(d) shows the cross sections of (b) a silicon strip waveguide with the waveguide width, $W_{si}$ = 450 nm, and height, $H_{si}$ = 250 nm, (across A-A″ cut), (c) a hybrid silicon-gold taper (across B-B″ cut), and (d) a plasmonic slot waveguide with the metal height, $H_{Au}$ = 250 nm, and the slot width, $W_{SU8}$ = 250 nm, (across C-C″ cut). The SU8 thickness ($T_{SU8}$) above the metal is 1.5 $\mu$m.

The coupling efficiency of the dielectric-to-plasmonic slot waveguide mode converter is studied and optimized with 3-D finite-difference time-domain (FDTD) simulations (Rsoft). Perfectly matched layer absorbing boundary conditions are used at all boundaries of the simulation domain. A mesh size of 20 nm is used in the dielectric waveguide regions, while an extra fine mesh (∼1 nm) is used to resolve the rapid variation of fields at the metal-dielectric interfaces. The finite element method (FEM) mode solver (Photon Design) is used to evaluate the effective indices of the mode converter and the waveguides. At the operating wavelength of 1550 nm, the material refractive indices of silicon (Si), $SiO_2$, SU8, and gold (Au) are 3.476, 1.45, 1.575, and $0.55 + i11.5$ [16], respectively. The coupling efficiency of mode converter is defined as the ratio of the power coupled into the plasmonic slot waveguide mode to the input power in the silicon waveguide. In order to calculate the coupling efficiency without the propagation loss of plasmonic slot waveguide, power coupled into the plasmonic slot waveguide mode is estimated by monitoring the output of the plasmonic slot waveguide and subtracting the propagation loss of plasmonic slot waveguide. In addition, the propagation loss of plasmonic slot waveguide is calculated by putting several monitors along the plasmonic slot waveguide and then fitting these transmissions, which are detected at the different positions of the plasmonic slot waveguide.

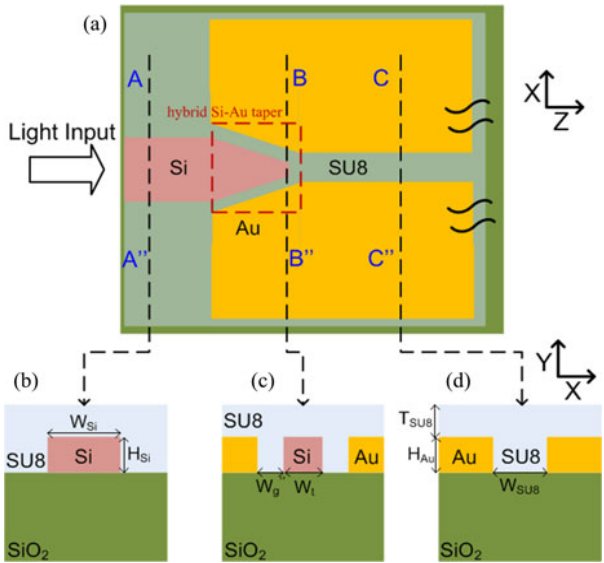

Fig. 1. Schematic diagrams of dielectric-to-plasmonic slot waveguide mode converter based on a hybrid silicon-gold taper. (a) Top view of the mode converter with a silicon strip waveguide at the input and a plasmonic metal slot waveguide at the output. Cross sections of (b) a silicon strip waveguide, (c) a hybrid silicon-gold taper, and (d) a plasmonic slot waveguide.

Mode matching, effective index matching and metal absorption in this structure and the tradeoffs among them for maximizing the overall efficiency are discussed below.

*A. Mode Matching*

Although it has been demonstrated by 2-D simulation that directly connecting a slab waveguide to a plasmonic slot waveguide could give a 70% coupling efficiency [2], a close match of the two optical fields is still the key to achieve a high conversion efficiency [12]. Fig. 2(a) shows the field distribution (Ex) in the *xz* plane using the 3-D FDTD simulation, which shows the evolution of the optical mode from the silicon strip waveguide to the plasmonic slot waveguide via a hybrid silicon-gold taper. Considering the practical fabrication limitations, the silicon taper tip is assumed to be 60 nm wide in the simulation. Fig. 2(b)–(f) show the cross-sections of fields at different locations along the mode converter, which are computed using the FEM mode solver. Fig. 2(b) and (f) show the field distributions of the silicon strip waveguide and the plasmonic slot waveguide, respectively. As it can be seen, the difference between the fields of the dielectric and plasmonic waveguides is evident. A good mode matching between these two waveguides through a low loss converter will therefore be important.

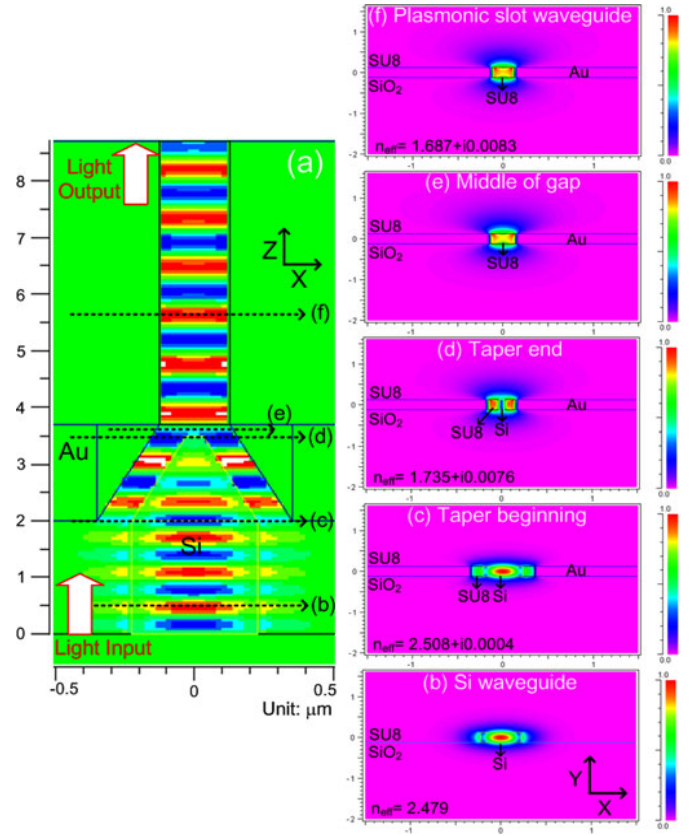

Fig. 2. Field distribution (Ex) of mode converter at 1550 nm. (a) Top view at the mode converter center. Cross-section views of (b) the silicon strip waveguide, (c) the hybrid silicon-gold taper at the taper beginning, (d) the hybrid silicon-gold taper at the taper end, (e) middle of gap between the silicon taper and the plasmonic slot waveguide, and (f) the plasmonic slot waveguide.

Fig. 2(c) and (d) show the field distributions at the beginning and at the end of the taper, respectively. The field distribution at taper beginning resembles a strip waveguide mode. The field distribution at the taper end becomes a hybrid plasmonic slot waveguide mode [15], which is confined in the gaps between silicon and gold. Lossless transformation from (c) to (d) is realized by narrowing the silicon waveguide tip to less than the diffraction limitation. As shown in the inset of Fig. 4, there is a gap (G) between the silicon taper and the plasmonic slot waveguide. Fig. 2(e) shows the mode at the middle of the gap. Fig. 2(f) shows the plasmonic slot waveguide mode, which could be strongly confined within the metallic slot. Through a metallic taper-funnel coupler, Fig. 2(e) and (f) should have similar field distributions. However, in order to achieve a high conversion efficiency, the mode matching at the input and output interfaces of the hybrid silicon-gold taper must be further improved including the interfaces of the silicon strip waveguide and the hybrid silicon-gold taper at taper beginning as well as the hybrid silicon-gold taper at taper end and plasmonic slot waveguide.

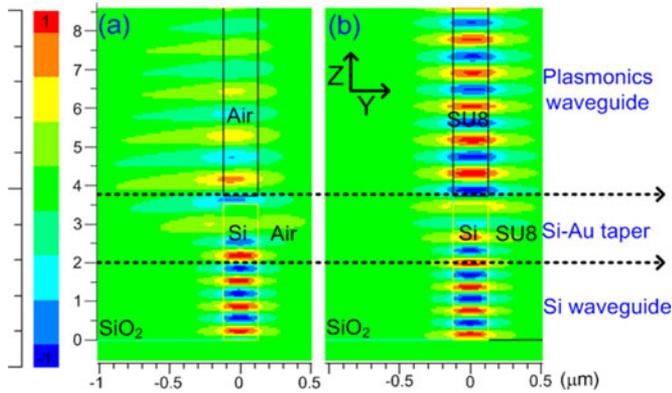

Fig. 3. Field distributions of mode converter at waveguide center (x = 0) using (a) air and (b) SU8 as a top cladding.

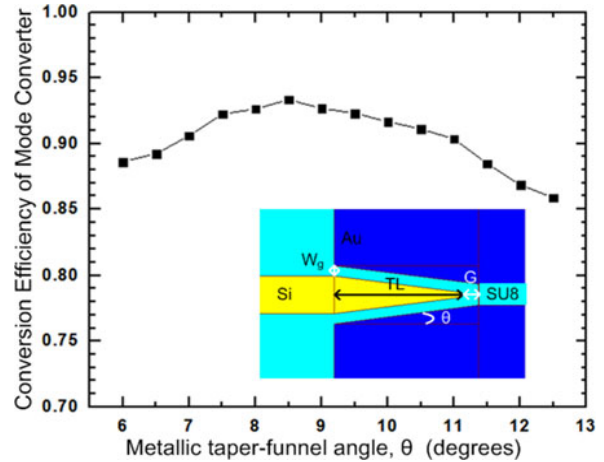

Fig. 4. Conversion efficiencies of the mode converter as a function of metallic taper-funnel angle ($\theta$). In this simulation, the angle of 7.5° and length of 1.5 $\mu$m for the silicon taper, G of 0.2 $\mu$m and the length of metallic taper-funnel coupler (1.7 $\mu$m) are fixed.

Before looking into mode matching at these interfaces, the alignment of the modes along the y (vertical) direction needs to be discussed in advance. In [13], the authors tuned the structural parameters of the taper (e. g. the taper angle, the tip width, and the distance between taper and plasmonic waveguide) and the metal gap to achieve a high theoretical coupling efficiencies of 88%. However, the mode matching along the $y$-direction is not investigated in their proposed structure. Due to asymmetric claddings along the $y$-direction, optical fields tend to shift to the higher index region. For example in [13], the authors use air as the top cladding material, the plasmonic slot waveguide mode within the hybrid silicon-gold taper and plasmonic slot waveguide will penetrate into the bottom cladding of $SiO_2$, causing an abrupt variation of the electromagnetic field, as shown in Fig. 3(a). The problem could be ameliorated by using materials with a higher refractive index as the top cladding.

In this paper, SU8, instead of air, is chosen to infiltrate the hybrid silicon-gold taper and plasmonic slot waveguide in order to ameliorate mode misalignment along the $y$-direction. As shown in Fig. 3(b), the plasmonic slot waveguide mode is "pulled up" to improve the mode matching with the silicon strip waveguide mode. Ideally, the modes should be perfectly aligned to achieve a high coupling efficiency, but in most cases misalignment along the $y$-direction is unavoidable due to the existence of refractive index asymmetry along the $y$-direction, which unfortunately reduces the conversion efficiency. For example, a silicon-plasmonic hybrid racetrack ring modulator was discussed in [5]. An electro-optic polymer as a top cladding having a higher refractive index than a bottom cladding of $SiO_2$ is filled in the slot of the MIM waveguide to achieve high-speed modulation. Therefore, in order to study the asymmetric cladding along the $y$-direction with the mode converter, we choose a prevalent polymer, SU8, as the top cladding of the mode converter.

Another benefit of using relatively higher refractive index materials as the top cladding is the reduction of the propagation loss of the plasmonic slot waveguide. In this case (using SU8 as the top cladding), a propagation loss as small as 0.32 dB/$\mu$m could be obtained due to its better guiding capability in the $y$-direction, especially at the SU8/$SiO_2$ interface. Another example of using air as the top cladding [13], a theoretical propagation loss of 1.5 dB/$\mu$m is encountered due to a poor guiding capability at the air/$SiO_2$ interface. Thus, the simulation discussed in the following texts uses SU8 as the top cladding.

As aforementioned, the field distributions in Fig. 2(b) and (c) have apparent differences due to the presence of metal in Fig. 2(c). Although most of the light is still confined in silicon, the evanescent wave interacts with metals, which alters the field distribution. Therefore, in order to achieve an effective mode matching, the gap between silicon and gold ($W_g$) needs to be large enough. In this case, we evaluate the conversion efficiencies of the mode converter as a function of metallic taper-funnel angle ($\theta$). $W_g$ would increase as the $\theta$ increases. The angle of 7.5° and length of 1.5 $\mu$m for the silicon taper, G of 0.2 $\mu$m and the length of metallic taper-funnel coupler (1.7 $\mu$m) are fixed in this simulation. The simulation result is shown in Fig. 4. The maximum conversion efficiency of hybrid silicon-gold taper is 93.3% and is obtained at $\theta = 8.5°$. The efficiency decreases when $\theta$ is larger than 8.5°. The reason is that a larger taper angle would cause a higher radiation loss. When $\theta$ is less than 8.5°, the efficiency decreases due to the poor mode matching.

Although Fig. 2(e) and (f) could have similar mode distribution via a metallic taper-funnel coupler, the longer metallic taper-funnel coupler will lead to higher metallic absorption loss. Therefore, as shown in the inlet of Fig. 4, the ratio of gap (G, which is between the silicon taper tip and the plasmonic slot waveguide.) and silicon taper length (TL) is studied for a high conversion efficiency. In this case, in order to maintain the same inclined angle for silicon taper, the width of silicon taper is appropriately widened relative to the gap. For the total length of gap adding silicon taper length (G + TL) being fixed as 1.7 $\mu$m, the high conversion efficiency of 93% is obtained when G is less than 0.2 $\mu$m. The conversion efficiency decreases as G increases beyond 0.2 $\mu$m. Finally, in order to facilitate the fabrication of the silicon taper with a larger taper width, the minimum taper width is 60 nm and its corresponding G of 0.2 $\mu$m for the hybrid silicon-gold taper are set for further optimization.

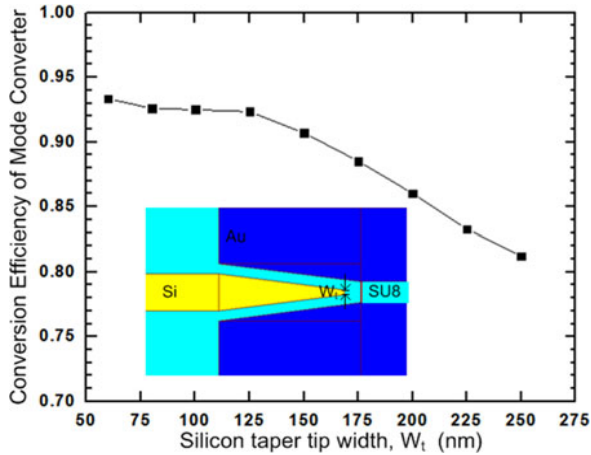

Fig. 5. Conversion efficiencies of the mode converter as a function of silicon taper tip width ($W_t$). In this simulation, the angle of 7.5° for the silicon taper, G of 0.2 $\mu$m, and the angle of 8.5° and length of 1.7 $\mu$m for the metallic taper-funnel coupler are fixed.

As shown in Fig. 2(d) and (f), two modes have difference in the field distribution because of the presence and absence of silicon, respectively. In this case, the conversion efficiency of the mode converter is studied by tuning the silicon taper tip width ($W_t$). In this simulation, the angle of 7.5° for the silicon taper, the G of 0.2 $\mu$m, and the angle of 8.5° and length of 1.7 $\mu$m for the metallic taper-funnel coupler are fixed. The simulation result is shown in Fig. 5. As expected the highest conversion efficiency of 93.3% is obtained for a narrow silicon taper tip width (60 nm in our case) and the conversion efficiency will decrease as the silicon taper tip width increases due to the poor mode matching. However, although a narrow taper tip width could enhance the mode matching, it would also cause a higher metallic absorption loss. For example, for an ideal case of taper tip width as 0 $\mu$m, the conversion efficiency of only 91.5% is obtained.

### B. Effective Index Matching

Effective index matching between the plasmonic slot waveguide and the hybrid silicon-gold taper is also studied. The effective index of the hybrid silicon-gold taper at the taper end can be engineered by tuning the gap between silicon and gold ($W_g''$) or the silicon taper tip width ($W_t$), as shown in the inset of Fig. 6. Fig. 6 shows the real and imaginary parts of effective indices of the hybrid silicon-gold taper at taper end as a function of $W_g''$. Both the real and imaginary parts of effective indices decreases with increasing $W_g''$. This suggests that a large $W_g''$ benefits from the effective index matching and also reduces the metallic absorption loss. In this simulation, $W_t$ is fixed at 60 nm in order to maintain the inclined angle of silicon taper.

The real part of the effective index of the plasmonic slot waveguide mode shown in Fig. 2(f) is 1.684. As shown in Fig. 6,

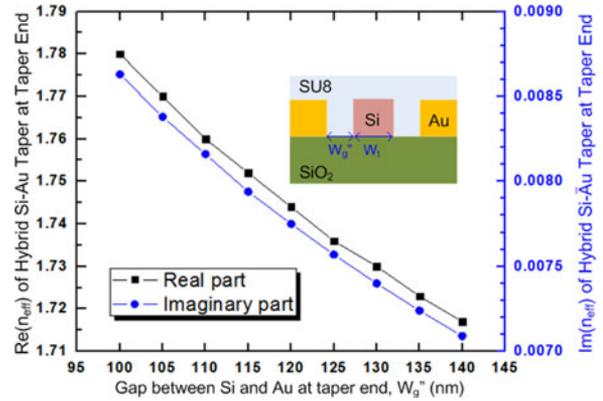

Fig. 6. Effective indices of a hybrid silicon-gold taper at taper end as a function of the gap ($W_g''$). $W_t$ of 60 nm is fixed in this simulation.

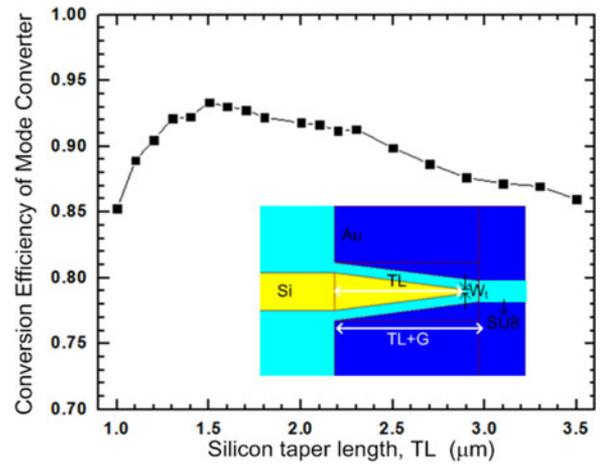

Fig. 7. Conversion efficiencies of the mode converter as a function of silicon TL. In this simulation, $W_t$ of 60 nm and G of 0.2 $\mu$m are fixed.

for the metallic taper-funnel angle of 8.5°, its corresponding real part of effective index is 1.735 ($W_g''$:125 nm). Considering the imperfect effective index matching between the hybrid silicon-gold taper at taper end and the plasmonic slot waveguide, these two modes have an index difference (real part) of about 0.05 caused by a higher refractive index of silicon. Although a lower effective index of hybrid silicon-gold taper could be obtained by increasing the $W_g''$, a larger $W_g''$ also increases metallic taper-funnel angle that causes a higher radiation loss.

### C. Metallic Absorption Loss within the Hybrid Si-Au Taper

Although a hybrid plasmonic slot waveguide mode can be generated through the hybrid silicon-gold taper, the longer hybrid silicon-gold taper the higher metallic absorption loss is. Therefore, the tradeoff between the radiation loss of the silicon taper and the metallic absorption loss within the hybrid silicon-gold taper is studied by optimizing the silicon TL. In this case, the length of the metallic taper-funnel coupler is varied as the TL is changed. Fig. 7 depicts the conversion efficiency of the mode converter as a function of silicon TL. The mode converter has high conversion efficiencies of more than 85% for different

TL from 1 to 3.5 $\mu$m. Finally, the highest conversion efficiency of 93.3% is achieved at a silicon TL of 1.5 $\mu$m. For the TLs less than 1.5 $\mu$m, the radiation loss of the silicon taper dominates the conversion loss; for the TL greater than 1.5 $\mu$m, the extra metallic absorption loss decreases the conversion efficiency.

## III. CONCLUSION

In this paper, through investigating the mode matching, the effective index matching, and the metallic absorption loss considerations, a hybrid silicon-gold taper with an overall length of 1.7 $\mu$m is studied and optimized. A conversion efficiency of 93.3% is achieved at 1550 nm. The simulation unveils that compromises have to be made between mode/index matching and absorption/radiation loss, and therefore, it is very challenging to achieve 100% conversion efficiency. A few design considerations are summarized below.

1) Considering the field distribution of hybrid silicon-gold taper at taper beginning, a small gap of $W_g$ would cause a poor mode matching with a silicon strip waveguide mode. $W_g$ shall therefore be large enough for the better mode matching. However, a large $W_g$ would generate a large $\theta$ so as to cause a higher radiation loss; the other mode mismatching is between hybrid silicon-gold taper at taper end and plasmonics slot waveguide. The presence and absence of silicon cause the field-distribution difference on these two modes. For a better mode matching, $W_t$ must be as small as possible. However, a smaller $W_t$ induces a higher metallic absorption loss;

2) For the effective index matching between the plasmonic slot waveguide and the hybrid silicon-gold taper at taper end, the $W_g''$ at taper end must be large enough to get a lower effective index to match the plasmonic slot waveguide. However, a larger $W_g''$ generates a large $\theta$ that causes a higher radiation loss. Hence, there is a tradeoff between radiation loss of metallic taper-funnel coupler and efficiently effective index matching;

3) The silicon strip waveguide mode can be converted to plasmonic slot waveguide mode through a hybrid silicon-gold taper. It indicates the metallic absorption loss within the hybrid silicon-gold taper could not be avoided. In order to reduce metallic loss, the gap between silicon and gold needs to increase due to the facts that the small size of plasmonic waveguide mode increases the propagation loss. The other approach to decrease the metallic loss is to reduce the TL. However, as shown in Fig. 7, the short TL increases the radiation loss.

Finally, considering the fabrication challenges for the hybrid silicon-gold taper that will decrease the coupling efficiency of this mode converter include (I) the imperfect alignment between the silicon taper and the plasmonic waveguide and (II) the surface roughness of metal pattern [13].

## ACKNOWLEDGMENT

C.-T. Chen would like to thank the Graduate Student Study Abroad Program sponsored by Ministry of Science and Technology, Taiwan to perform his advanced research at the Department of Electrical & Computer Engineering, University of Texas at Austin.

**Chin-Ta Chen** received the B.S. degree in electronic engineering from the National Changhua University of Education, Changhua City, Taiwan, and the M.S. degree in optical sciences from the National Central University, Jhongli, Taiwan, in 2008 and 2010, respectively, where he is currently working toward the Ph.D. degree in the Department of Optics and Photonics, National Central University.

His research interests include optical interconnect, silicon photonics, and radio-frequency design.

**Xiaochuan Xu** received the B.Sc. and M.S. degrees in electrical engineering from the Harbin Institute of Technology, Harbin, China, in 2006 and 2009, respectively, and the Ph.D. degree in electrical and computer engineering from the University of Texas, Austin, TX, USA, in 2013.

His research interests include flexible photonics, nonlinear optics, fiber optics, and silicon photonics

**Amir Hosseini** (S'05–M'13) received the B.Sc. degree in electrical engineering from the Sharif University of Technology, Tehran, Iran, in 2005, the M.Sc. degree in electrical and computer engineering from Rice University, Houston, TX, USA, in 2007, and the Ph.D. degree in electrical and computer engineering from the University of Texas at Austin, Austin, TX, in 2011.

He has been involved in research on modeling, design, fabrication, and characterization of optical phased array technology, true-time delay lines, and high performance optical modulators. He is a Prince of Wales' scholar in 2011 and received the Ben Streetman Award in 2012, and has authored or coauthored more than 100 peer reviewed technical papers. He is a member of the Optical Society of America and International Society for Optical Engineers. He has been serving as the Principal Investigator for an Air Force Research Laboratory sponsored project on polymer optical modulators since 2012.

**Zeyu Pan** (S'12) received the B.S. degree in optical information science and technology from the Nanjing University of Science and Technology, in 2010, and the M.S. degree in electrical and computer engineering from the University of Alabama in Huntsville, Huntsville, AL, USA, in 2013. He is currently working toward the Ph.D. degree in the University of Texas at Austin, Austin, TX, USA. He has authored and coauthored more than 20 journal and conference papers. His research interests include the polymer waveguide, printed flexible electronics and photonics, patch array antennas, thermo- and electrooptic polymers, phase delay, nanofabrication, RF measurements, inkjet printing, imprinting, and numerical algorithms.

**Harish Subbaraman** (M'09) received the M.S. and Ph.D. degrees in electrical engineering from the University of Texas at Austin, Austin, TX, USA, in 2006 and 2009, respectively. With a strong background in RF photonics and X-band Phased Array Antennas, he has been working on optical true-time-delay feed networks for phased array antennas for the past seven years. Throughout these years, he has laid a solid foundation in both theory and experimental skills. His current research interests include printing and silicon nanomembrane-based flexible electronic and photonic devices, polymer photonics, slow-light photonic crystal waveguides, carbon nanotube, and silicon nanoparticle nanofilm-based ink-jet printed flexible electronics, and RF photonics. He has served as a PI on eight SBIR/STTR Phase I/II projects from National Aeronautics and Space Administration, Air Force, and Navy. He has more than 70 publications in refereed journals and conferences.

**Xingyu Zhang** (S'13) received the B.S. degree in electrical engineering from the Beijing Institute of Technology, Beijing, China, in 2009, and the M.S. degree in electrical engineering from the University of Michigan, Ann Arbor, MI, USA, in 2010. He is currently working toward the Ph.D degree at the University of Texas, Austin, TX, USA.

His current research interests include design, fabrication, and characterization of silicon and polymer hybrid integrated microwave nanophotonic devices used in optical interconnects and RF sensing, including highly linear broadband optical modulators, high-speed subvolt low-dispersion slot photonic crystal waveguide modulators, sensitive integrated photonic electromagnetic field sensors, highly efficient strip-to-slot mode converters, and broadband integrated antennas. He has published as the first author about 20 peer-reviewed papers in journals and conferences during the Ph.D. degree, including one invited paper in IEEE journal of selected topics in quantum electronics.

Mr. Zhang has so far served as Reviewer for 13 prestigious journals in his research area. He received the Engineering Scholarship Award at UT-Austin and Full Financial Fellowship at UM-Ann Arbor. He has industry internship experiences in IBM and Oracle. He is the Student Member of the International Society for Optical Engineers (SPIE) and *Optical Society of America*.

**Ray T. Chen** (M'91–SM'98–F'04) received the B.S. degree in physics from National Tsing-Hua University in 1980, Hsinchu City, Taiwan, and the M.S. degree in physics in 1983 and the Ph.D. degree in electrical engineering in 1988, both from the University of California, Oakland, CA, USA. He holds the Cullen Trust for Higher Education Endowed Professorship at University of Texas at Austin, TX, USA and is the Director of Nanophotonics and Optical Interconnects Research Lab within the Microelectronics Research Center. He is also the Director of a newly formed AFOSR MURI-Center involved faculty from Stanford, University of Illinois at Urbana-Champaign, Rutgers for Silicon Nanomembrane. He joined UT Austin as a Faculty to start optical interconnect research program in the Electrical and Computer Engineering Department in 1992. Prior to his UT's professorship, he was a Research Scientist, Manager, and the Director of the Department of Electro-optic Engineering in Physical Optics Corporation in Torrance, Torrance, CA, USA, from 1988 to 1992.

He also served as the CTO/Founder and the Chairman of the board of Radiant Research from 2000 to 2001, where he raised 18 million dollars A-Round funding to commercialize polymer-based photonic devices. He also serves as the Founder and the Chairman of the board of Omega Optics Inc. since its initiation in 2001. More than 5 million dollars of research funds were raised for Omega Optics. His research work has received with 110 research grants and contracts from such sponsors as Department of Defense, National Science Foundation, Department of Energy, National Aeronautics and Space Administration, National Institutes of Health, Environmental Protection Agency, the State of Texas, and private industry. The research topics are focused on three main subjects: 1) Nanophotonic passive and active devices for optical interconnect and biosensing applications, 2) Polymer-based guided-wave optical interconnection and packaging, and 3) True time delay wide band-phased array antenna. Experiences garnered through these programs in polymeric material processing and device integration are pivotal elements for the research work conducted by his group.

Dr. Chen's group at UT Austin has reported its research findings in more than 650 published papers including more than 85 invited papers. He holds 20 issued patents. He has Chaired or been a program-committee member for more than 90 domestic and international conferences organized by the IEEE, The International Society of Optical Engineering (SPIE), Optical Society of America, and PSC. He has served as an editor, coeditor, or coauthor for 22 books. He has also served as a Consultant for various federal agencies and private companies and delivered numerous invited talks to professional societies. He is a Fellow of the OSA and SPIE. He received the 1987 UC Regent's dissertation fellowship and of 1999 UT Engineering Foundation Faculty Award for his contributions in research, teaching, and services. He received IEEE Teaching Award in 2008. Back to his undergraduate years in National Tsing-Hua University, he led a university debate team in 1979 that received the national championship of national debate contest in Taiwan. Forty-four students have received the EE Ph.D. degree in his research group at UT Austin.